\documentclass[aip,jcp,twocolumn]{revtex4}

\usepackage[version=3]{mhchem} 

\usepackage{amsmath,amssymb}
\usepackage{graphicx,color,subfigure}
\usepackage{amsfonts}
\usepackage{comment}
\usepackage{placeins}
\usepackage{multirow}
\usepackage{diagbox}

\usepackage{balance}
\usepackage{xspace}
\usepackage{algorithm,algorithmic}

\usepackage{appendix}

\usepackage{epsfig,amssymb,amsmath,amsthm,amsfonts,amsbsy,mathrsfs}
\usepackage{graphicx,color}
\usepackage{amsmath}
\usepackage{amssymb}
\usepackage{epstopdf}

\usepackage{graphicx} 
\usepackage{lastpage}
\usepackage[format=plain,justification=raggedright,singlelinecheck=false,font=small,labelfont=bf,labelsep=space]{caption}

\allowdisplaybreaks

\usepackage[version=3]{mhchem} 

\newcommand{\bt}{\boldsymbol{\theta}}

\begin{document}

\title{Reducing circuit depth in adaptive variational quantum algorithms via effective Hamiltonian theories}

\author{Jie Liu}
\affiliation{Hefei National Laboratory for Physical Sciences at the Microscale, University of Science and Technology of China, Hefei, Anhui 230026, China}

\author{Zhenyu Li}
\email{zyli@ustc.edu.cn (Zhenyu Li)}
\affiliation{Hefei National Laboratory for Physical Sciences at the Microscale, University of Science and Technology of China, Hefei, Anhui 230026, China}

\author{Jinlong Yang}
\email{jlyang@ustc.edu.cn (Jinlong Yang)}
\affiliation{Hefei National Laboratory for Physical Sciences at the Microscale, University of Science and Technology of China, Hefei, Anhui 230026, China}

\begin{abstract}
Electronic structure simulation is an anticipated application for quantum computers. Due to high-dimensional quantum entanglement in strongly correlated systems, the quantum resources required to perform such simulations are far beyond the capacity of current quantum devices. To reduce the quantum circuit complexity, it has been suggested to incorporate a part of the electronic correlation into an effective Hamiltonian, which is often obtained from a similarity transformation of the electronic Hamiltonian. In this work, we introduce a new transformation in the form of a product of a linear combination of excitation operators to construct the effective Hamiltonian with finite terms. To demonstrate its accuracy, we also consider an equivalent adaptive variational algorithm with this transformation and show that it can obtain an accurate ground state wave function. The effective Hamiltonian defined with this new transformation is incorporated into the adaptive variational quantum algorithms to maintain constant-size quantum circuits. The new computational scheme is assessed by performing numerical simulations for small molecules. Chemical accuracy is achieved with a much shallower circuit depth.
\end{abstract}

\maketitle
\section{Introduction}
Rapid advances in quantum computing technologies have stimulated the development of novel quantum algorithms for electronic structure simulations. The variational quantum eigensolver (VQE) is considered the most feasible scheme for experimental and theoretical realization of quantum simulation of electronic structures on near-term quantum devices~\cite{AspDutLov05,DuXuPen10,PerMcCSha14,MalBabKiv16,KanMezTem17,HemMaiRom18,AruAryBab20,Pre18,CaoRomOls19,McAEndAsp20,ZhoWanDen20}. As a hybrid quantum-classical algorithm, the VQE obtains the statistical energy expectation value estimate by repeating state preparation and measurement many times on quantum computers and then feeds the total energy back to classical computers to optimize variational parameters~\cite{PerMcCSha14,McCRomBab16,RomBabMcC18}.

The wave function ansatz for state preparation,
\begin{equation}\label{eq:wave}
    |\Psi(\bt)\rangle = \hat{U}(\bt) |\Psi_\mathrm{ref}\rangle,
\end{equation}
determines the accuracy and quantum circuit complexity of a VQE algorithm. Here, $|\Psi_\mathrm{ref}\rangle$ is the reference state that can be easily prepared on quantum computers and $\hat{U}(\bt)$ is a parametrized unitary transformation that can be implemented on quantum computers in polynomial time. Unitary coupled cluster (UCC) is one of the most widely used wave function ansatz within the VQE framework~\cite{PerMcCSha14,SheZhaZha17,RomBabMcC18,LeeHugHea19}. The UCC wave function is in principle exact by including all possible excitations, while this comes with a prohibitive computational cost as the system size increases. In practice, a UCC with single and double excitations (UCCSD) is often used for efficient implementation but it fails to accurately capture strong electronic correlation effects.~\cite{LeeHugHea19,EvaChaScu19,LiuWanLi20}

To reduce the circuit complexity, many recent contributions have been devoted to adaptive VQE algorithms.~\cite{GriEcoBar19,TanShkBar21,YorArmBar20,LiuLiYan21} Adaptive Derivative-Assembled Pseudo-Trotter (ADAPT) ansatz~\cite{GriEcoBar19} VQE iteratively builds a system-adapted wave function as a product of general one- and two-body exponential operators. A qubit version of ADAPT-VQE further compresses circuit depth by replacing the fermionic excitation operators with corresponding Pauli-string operators that include only Pauli X and Y.~\cite{TanShkBar21} Analogously, the qubit-excitation-based VQE (QEB-VQE) parameterizes the wave function with the one- and two-body qubit-excitation operators, in which multiqubit operators with a maximum length of 4 are involved in the operator pool~\cite{YorArmBar20}. In addition, qubit coupled cluster singles and doubles (QCCSD) VQE, a qubit version of UCCSD-VQE, has been proposed based on the particle preserving exchange gate to achieve qubit excitations~\cite{XiaKai20}. However, due to highly entangled electron-electron interactions in strongly correlated systems, quantum circuits for representing the exact electronic wave function are still too complicated to implement on near-term quantum devices.

Inspired by effective Hamiltonian theories,~\cite{Fre74,Bra77,TauBar06,BarMus07,Eva14,LiEva17} one can include a part of electron correlation in the effective Hamiltonian
\begin{equation}
    \hat{H}^\prime(\bt) = \hat{D}^\dag(\bt) \hat{H} \hat{D}(\bt)
\end{equation}
and leave the rest to a correlated electronic wave function $|\Psi\rangle$. Here, $\hat{D}^\dag(\bt)$ is a parametrized transformation, often in an exponential form, and the effective Hamiltonian employs a Hermitian form to accommodate the variational nature of the VQE. While effective Hamiltonian theories, such as coupled cluster theory~\cite{BarMus07} and the similarity renormalization group approach,~\cite{Eva14} have a long history in electronic structures simulation, they were introduced in quantum computational chemistry until very recently.~\cite{BauBylKri19,RyaLanGen20,MotGujRic20,McATew20,LanRyaIzm21} With an appropriate partition of electron correlation, one can in principle control quantum circuit depth in trade with the Hamiltonian size. Therefore, the exploitation of effective Hamiltonian theories within the VQE framework provides a potential way to overcome the limitations of contemporary quantum computers.

Effective Hamiltonian theories have been used in quantum computational chemistry to reduce the number of qubits required. Due to the immature quantum computation platform, most quantum simulations of electronic structures are limited to minimal basis sets or small active space. To improve the accuracy of basis sets for limited-qubit VQE simulations, one common strain is to include the dynamical electron correlation effects in a similarity-transformed Hamiltonian and leave the static correlation to the wave function defined in a space much smaller than the one spanned by a large basis set and many electrons. Motta et al. incorporated the canonical transcorrelated F12 Hamiltonian into the UCCSD-VQE algorithm to suppress the Coulomb singularity of the electronic interaction and achieve a larger basis set quality for explicitly evaluating correlation energies.~\cite{MotGujRic20} Instead of a unitary transformation used in Ref.~\citenum{MotGujRic20}, McArdle et al. employed a nonunitary transformation in Jastrow form, and as a consequence, the imaginary time evolution approach was introduced to overcome the limitation of violating the variational principle.~\cite{McATew20} Another strain of such works is the double unitary coupled cluster formalism, which decouples dynamical and static correlation effects into the effective Hamiltonian and the corresponding (CAS) eigenvalue problem.~\cite{BauBylKri19}

Alternatively, one can construct an efficient Hamiltonian to reduce the quantum circuit depth. An iterative version of the qubit coupled cluster (iQCC) method employs an effective Hamiltonian ''dressed'' by a series of canonical transformations to maintain constant-size quantum circuits.~\cite{RyaLanGen20} Unlike the ADAPT-VQE and QEB-VQE methods, the iQCC method constructs canonical transformations in the form of exponential unitary operators directly in the multiqubit space. A unitary transformation with an exact quadratic truncation of the Baker-Campbell-Hausdorff (BCH) expansion was constructed as the exponent of an involuntary linear combination of anticommuting Pauli products.~\cite{LanRyaIzm21} As a consequence, it achieves a significant improvement over the original version of iQCC.

In this work, we consider the construction of an effective Hamiltonian with a sequence of linear transformations
\begin{equation}
    \hat{D}(\bt) = \prod_m (1+\theta_{m,1}\tau_{m,1}+\cdots),
\end{equation}
where $\tau$ are anti-Hermitian operators. These transformations can be regarded as the first-order Taylor-expansion (FT) approximation to corresponding unitary transformations in the exponential form. Equivalently, we can define an adaptive variational algorithm with these transformation
\begin{equation}
    |\Psi(\bt)\rangle = \hat{D}(\bt) |\Psi_\mathrm{ref}\rangle,
\end{equation}
termed Adaptive Derivative-Assembled First-Order Taylor-Expansion (ADAFT) ansatz. Analogous to the ADAPT ansatz, the ADAFT wave function is iteratively constructed. As demonstrated in Sec.~\ref{sec:theory1}, the ADAFT ansatz holds the same convergence condition as the ADAPT ansatz even though it is considered the first order approximation to the ADAPT ansatz. Given the close connection between the ADAPT and ADAFT ansatz, we employ the ADAPT ansatz for state preparation and the FT transformation to {\it dress} the Hamiltonian. We name this method ADAPT-FT. In contrast to the iQCC method, the ADAPT-FT method can be applied to (spin-adapted) fermionic operators, qubit excitation operators or any Pauli string operators. In addition, since the FT transformation is a linear combination of excitation operators, we can include in principle as many operators as we want in each iteration to control the growth factor of the effective Hamiltonian.

The rest of this paper is organized as follows. In Section~\ref{sec:theory1}, we briefly review adaptive VQE algorithms, including ADAPT-VQE, qubit-ADAPT-VQE and QEB-VQE algorithms, and then we introduce the ADAFT ansatz and compare it to the ADAPT ansatz. The scheme for transforming the Hamiltonian and the ADAPT-FT method are introduced in Section~\ref{sec:theory2}. In Section~\ref{sec:results}, we first assess the convergence and accuracy of the ADAFT ansatz and the performance of the ADAPT-FT method is then discussed. The conclusion is given in Section~\ref{sec:conclusion}.

\section{Theory}\label{sec:theory}
The general Hamiltonian of a many-electron system is expressed in the second-quantized as
\begin{equation}\label{eq:Hamiltonian}
    \hat{H} = \sum_{pq}^{N_{so}} h^p_q a^\dag_p a_q + \frac{1}{2} \sum_{pqrs}^{N_{so}} v^{pq}_{sr}  a^\dag_p a^\dag_q a_r a_s
\end{equation}
where $h^p_{q}$ are the one-electron integrals, including kinetic energy and ionic potential, and $v^{pq}_{rs}$ is the two-electron repulsion integral. $N_{so}$ is the number of molecular spin-orbitals. $a_p^\dag$ and $a_p$ are the second-quantized creation and annihilation operators that satisfy the anticommutation relations
\begin{equation}
    \{a_p^\dag, a_q\}=\delta_{pq},\ \{a_p^\dag,a_q^\dag\}=\{a_p,a_q\}=0.
\end{equation}
The time-independent Schr\"odinger equation is
\begin{equation}
    \hat{H}|\Psi\rangle = E |\Psi\rangle.
\end{equation}
In the quantum computing context, the solution of the Schr\"odinger equation is often represented as Eq.~\ref{eq:wave}. The unitary transformation $\hat{U}$ can be either an exponential cluster operator, e.g., UCCSD and UCCGSD, in the form of
\begin{equation}
    \hat{U} = e^{\hat{X}}
\end{equation}
or a product of exponential operators, e.g. iQCC,~\cite{RyaLanGen20} $k$-UpCCGSD~\cite{LeeHugHea19} and ADAPT-VQE~\cite{GriEcoBar19}, in the form of
\begin{equation}
    \hat{U} = e^{\theta_k \hat{X}_k} \cdots e^{\theta_1 \hat{X}_1}.
\end{equation}
The latter form often generates a more compact wave function ansatz when combined with an iterative VQE procedure. In this work, we focus on these adaptive variational algorithms, which build the wave function ansatz by iteratively growing the unitary transformation that intends to restore the largest amount of the correlation energy.

\subsection{Adaptive VQE algorithms} \label{sec:theory1}
In adaptive VQE algorithms, the wave function is iteratively updated with
\begin{equation}\label{eq:wave_adapt}
    | \Psi_k(\bt) \rangle =  e^{\theta_k \tau_k} |\Psi_{k-1} (\bt) \rangle
\end{equation}
where $|\Psi(0)\rangle = |\Psi_0\rangle$ is the reference state. To guarantee that the transformation is unitary, we assume $\tau_k$ to be an anti-Hermitian operator selected from a predefined operator pool. The energy functional in the $k$-th iteration is minimized by
\begin{equation}\label{eq:energy}
    E_k = \min_{\bt}  \langle \Psi_k(\bt) | \hat{H} | \Psi_k(\bt) \rangle.
\end{equation}
The analytical gradient of the energy functional with respect to parameters $\{\theta_l\}_{1}^{k}$ is
formulated as
\begin{equation}\label{eq:gradient}
\begin{split}
    G_{l} &= \frac{\partial E_k}{\partial \theta_l} \\
    &= \langle \Psi_k(\bt) | \hat{H} \prod_{m=l+1}^{k} e^{\theta_m \tau_m} \tau_l \prod_{n=1}^l e^{\theta_n \tau_n} |\Psi_0 \rangle  \\
    &- \langle \Psi_0 | \prod_{m=l}^{1} e^{-\theta_m \tau_m} \tau_l \prod_{n=k}^{l+1} e^{-\theta_n \tau_n} \hat{H}  |\Psi_k(\bt) \rangle
    \end{split}
\end{equation}
The convergence of the wave function can be assessed by the residual gradient
\begin{equation}\label{eq:acse_res}
\begin{split}
    R_u &= G_{k+1}|_{\theta_{k+1}=0,\tau_{k+1}=\tau_u} \\
    &= \langle \Psi_k| [\hat{H},\tau_u] |\Psi_k \rangle.
    \end{split}
\end{equation}
Hence, the convergence condition is defined as
\begin{equation}\label{eq:conv}
    |\mathbf{R}| = \sqrt{\sum_u |R_u|^2} < \epsilon
\end{equation}

\subsubsection{Fermionic Operators}
Grimsley {\it et al.}~\cite{GriEcoBar19} proposed that the exact wave function can be expressed as an arbitrarily long product of general one- and two-body exponentiated operators with $\{\tau_u\}$ being anti-Hermitian operators
\begin{equation}\label{eq:operator}
    \begin{split}
        &\tau^p_q=a_p^\dag a_q - a_q^\dag a_p \\
        &\tau^{pq}_{rs} = a_p^\dag a_q^\dag a_r a_s - a_s^\dag a_r^\dag a_q a_p.
    \end{split}
\end{equation}
Note that $R_u=0$ are exactly the anti-Hermitian contracted Schr\"odinger equations (ACSE) presented in Ref.~\citenum{Maz06a}, namely, the wave function of Eq.~\ref{eq:wave_adapt} at convergence is a solution of the ACSE up to the accuracy of $\epsilon$. The ADAPT-VQE establishes a well-defined order for a sequence of iteratively determined exponential operators, which avoids errors from Trotterization in the UCC ansatz.

Smart and Mazziotti recently proposed a quantum solver of the contracted Schr\"odinger equations (CSE) and demonstrated it on both a quantum simulator and two IBM quantum processing units\cite{SmaMaz21}. In this quantum solver, the trial wave function for approximating the solution of the CSE is updated by
\begin{equation}\label{eq:wf_acse}
     |\Psi_{k+1}\rangle = e^{\theta_k \hat{X}_k} |\Psi_k\rangle,
\end{equation}
where $\hat{X}_k=\sum_u \theta_k^{pq,rs} \tau_{rs}^{pq}$ is restricted to be anti-Hermitian and the coefficients $\theta_k^u$ are determined as
\begin{equation}\label{eq:acse_coef}
     \theta^{pq;rs}_k = \langle \Psi_k | [a^\dag_p a^\dag_q a_r a_s, \hat{H}] | \Psi_k \rangle
\end{equation}
The iterative minimization procedure stops until the energy or the residuals of the ACSE cease to decrease. While this method has fewer parameters than the ADAPT method, the wave function ansatz of Eq.~\ref{eq:wf_acse} may result in more complicated quantum circuits at convergence because the cluster operator that includes many excitation operators is used in the unitary transformation.

\subsubsection{Qubit Excitation Operators}
The general one- and two-body operators can be mapped to qubit operators using Jordon-Wigner (JW) encoding methods, for $p < q < r < s$,
\begin{equation}\label{eq:operator_qubit}
\begin{split}
        \tau^p_q &=  (Q^\dag_p Q_q - Q^\dag_q Q_p) \prod_{t=p+i}^{q-1} Z_t \\
        \tau^{pq}_{rs} &= (Q_p^\dag Q_q^\dag Q_r Q_s - Q_s^\dag Q_r^\dag Q_q Q_p) \prod_{t=p+1}^{q-1} Z_t \prod_{v=r+1}^{s-1} Z_v,
\end{split}
\end{equation}
with
\begin{equation}
\begin{split}
    Q_p^\dag &= \frac{1}{2}(X_p-iY_p) \\
    Q_p &= \frac{1}{2}(X_p+iY_p)
\end{split}
\end{equation}
Here, $Q^\dag_p$ and $Q_p$ are qubit creation and annihilation operators, respectively. $\{X,Y,Z\}$ are the Pauli operators. The general single and double qubit-excitation operators are defined as~\cite{YorArmBar20}
\begin{equation}\label{eq:qubit_excitation}
    \begin{split}
        &\tilde{\tau}^p_q=Q_p^\dag Q_q - Q_q^\dag Q_p \\
        &\tilde{\tau}^{pq}_{rs} = Q_p^\dag Q_q^\dag Q_r Q_s - Q_s^\dag Q_r^\dag Q_q Q_p.
    \end{split}
\end{equation}
The Pauli-Z chains responsible for the fermionic anticommutation relation have been removed from Eq.~\eqref{eq:operator_qubit}.

\subsubsection{Pauli-String Operators}
Analogous to the QEB-VQE, the qubit-ADAPT-VQE generates the operator pool from multiqubit operators with a maximum length of 4, while it discards the physical structure and employs a more general form of Pauli strings. Decomposing the operators in Eq.~\ref{eq:qubit_excitation} into elemental Pauli strings, we obtain
\begin{equation}
    \begin{split}
        &\bar{\tau}^p_q \in \{iX_pY_q,iY_pX_q\}\\
        &\bar{\tau}^{pq}_{sr} \in\{iX_pY_qX_rX_s,iY_pX_qX_rX_s,iY_pY_qY_rX_s,iY_pY_qX_rY_s,\\
        &\quad iX_pX_qY_rX_s,iX_pX_qX_rY_s,
        iY_pX_qY_rY_s,iX_pY_qY_rY_s\}.
    \end{split}
\end{equation}
It is clear that only Pauli strings with an odd number of Y's exist to guarantee that operators are real. The iterative optimization procedure for qubit-ADAPT-VQE is exactly the same as that for ADAPT-VQE and QEB-VQE. It is worth mentioning that qubit-ADAPT-VQE shares the same operator pool as iQCC in the first iteration, while in the following iterations, iQCC generates a more general operator pool without the limitation of the Pauli string lengths.~\cite{RyaLanGen20} As such, some advanced techniques proposed in the iQCC, e.g., an efficient screening procedure~\cite{RyaLanGen20} and constructing an involutory linear combination of entangled operators~\cite{LanRyaIzm21}, are also applicable to reduce the quantum resource required in qubit-ADAPT-VQE.

\subsection{ADAFT Ansatz}\label{sec:theory2}
Nakatsuji suggested that the exponential operators could be expanded to the first order while the wave function remained exact.~\cite{Nak00} They called this method the iterative configuration interaction method. Mazziotti has recently expanded the wave function as~\cite{Maz20}
\begin{equation}\label{eq:DL}
    |\Psi\rangle = \prod_{i=1}^k (1+\lambda \hat{F}_\lambda) |\Psi_\mathrm{ref}\rangle
\end{equation}
with the two-body operator
\begin{equation}
    \hat{F}_\lambda = \sum_{pqrs} {}^2F^{pqrs}_\lambda a_p^\dag a_q^\dag a_r a_s.
\end{equation}
Although Eq.~\ref{eq:DL} implies the exact solution of the CSE, the number of terms in the $\hat{F}_\lambda$ operator scales as $\mathcal{O}(N^4)$. This presents a great challenge to implementing this method on near-term quantum devices as the system size increases.

Here, we introduce an adaptive procedure to construct the wave function ansatz as done in the ADAPT
\begin{equation}\label{eq:wave_taylor}
        | \Psi_k(\bt) \rangle =  (1 + \theta_k \tau_k) |\Psi_{k-1} (\bt) \rangle.
\end{equation}
The convergence condition is simply formulated as
\begin{equation}
    \begin{split}
    R_u &= G_{k+1}|_{\theta_{k+1}=0,\tau_{k+1}=\tau_u} \\
    &= \langle \Psi_k| [\hat{H},\tau_u] |\Psi_k \rangle < \epsilon
    \end{split}
\end{equation}
where we assume that the wave function $\Psi_k$ is normalized. It is clear that this convergence condition is exactly the same as that in the ADAPT ansatz. Note that the expansion of the exponential operators is no longer unitary. Therefore, the wave function of Eq.~\eqref{eq:wave_taylor} is only size extensive in its energy when the wave function converges to the exact ground (or excited) state.

Considering that each iteration will add one time more configurations to the updated wave function in the form of Eq.~\ref{eq:wave_taylor}, the number of configurations increases as $2^k$. An alternative strategy to reduce the number of iterations is to increase the number of operators to be updated in each iteration, namely,
\begin{equation}\label{eq:wave_taylor_n}
    | \Psi_k(\bt) \rangle =  (1 + \sum_u^d \theta_{k,u} \tau_{k,u}) |\Psi_{k-1} (\bt) \rangle.
\end{equation}
It is easy to demonstrate that the residual gradients of Eq.~\eqref{eq:wave_taylor_n} have exactly the same form as that of Eq.~\eqref{eq:wave_taylor}. Although the ADAFT method using Eq.~\ref{eq:wave_taylor_n} results in a more rapid increase in the dimension of the wave function, a much faster convergence can be achieved as discussed in Ref.~\cite{LiuLiYan21}.

\subsection{Effective Hamiltonian Theories} \label{sec:theory3}
Analogous to the adaptive variational algorithm, a unitary transformation for constructing the effective Hamiltonian can be defined as
\begin{equation}\label{eq:transform_unitary}
    \hat{D}_m = e^{\theta_m \tau_m} \hat{D}_{m-1}.
\end{equation}
However, this unitary transformation results in an infinite BCH expansion of the effective Hamiltonian when anti-Hermitian fermionic or qubit excitation operators are used. Alternatively, we employ a linear combination form of these operators to construct the transformation
\begin{equation}\label{eq:transform_taylor}
    \hat{D}_m = (1 + \theta_m \tau_m) \hat{D}_{m-1}
\end{equation}
or
\begin{equation}\label{eq:transform_taylor_n}
    \hat{D}_m = (1 + \sum_u^d \theta_{m,u} \tau_{m,u}) \hat{D}_{m-1}.
\end{equation}
The energy is defined as
\begin{equation}\label{eq:dressedH}
    E_{k,m} = \langle \Psi_k |\hat{H}^\prime_m | \Psi_k \rangle
\end{equation}
with the effective Hamiltonian
\begin{equation}\label{eq:dressed_hamiltonian}
    \hat{H}^\prime_m = \hat{D}_m^\dag \hat{H} \hat{D}_m
\end{equation}
To minimize the quantum hardware demand for state preparation of $|\Psi_k\rangle$, $k$ can be set to be as small as possible. However, a simple wave function ansatz will result in a fast growth of the effective Hamiltonian. Finding a delicate balance between the wave function ansatz and the transformed Hamiltonian is crucial for a practical implementation of this algorithm.

Overall, the ADAPT-FT procedure is described as follows:
\begin{itemize}
    \item Define the operator pool.
    \item Prepare the correlated wave function with adaptive VQE algorithms. Here, the number of variational parameters in unitary transformations, $k$, is fixed.
    \item Construct the effective Hamiltonian with Eq.~\eqref{eq:dressed_hamiltonian}, in which operators with the largest residual gradients are included.
    \item Update the transformation $D$ and return to Step 3.
    \item If the maximal number of iterations or convergence is reached, exit.
\end{itemize}

\section{Results}\label{sec:results}
In this section, we numerically study the convergence and accuracy of the ADAFT method and then assess its performance. All calculations are performed with the in-house Python code, which uses OpenFermion~\cite{openfermion} for mapping fermionic operators onto qubit operators and PySCF~\cite{pyscf} for all one- and two-electron integrals. The energy and wave function are optimized with the Broyden-Fletcher-Goldfarb-Shannon (BFGS) algorithm implemented in SciPy~\cite{scipy}. Gradients are computed with the analytical approach. Reference results from complete active space configuration interaction (CASCI) calculations are used for comparison. The convergence criterion is the variance of the expectation value of the Hamiltonian,
\begin{equation}\label{eq:conv_H}
     \delta H = \sqrt{\langle \Psi | \hat{H}^2 |\Psi\rangle - \langle \Psi | \hat{H} | \Psi \rangle ^2} < \epsilon.
\end{equation}
A spin-penalized Hamiltonian in the form of,
\begin{equation}
    \hat{H}_s = \Hat{H} + \frac{\mu}{2} \hat{S}^2,
\end{equation}
is used to guarantee that the ground state converges to the singlet state. $\hat{S}$ is the spin operator and $\mu=0.5$ Hartree.

To ensure the consistency of method naming, instead of ADAPT, QEB and qubit-ADAPT, we label these adaptive wave function ansatz as $x$-ADAPT, in which the prefix $x$ being $f$, $q$ and $p$ indicates fermionic, qubit excitation and general Pauli-string operators, respectively. Correspondingly, $x$-ADAFT indicates the ADAFT ansatz with different types of operator pools. Calculations are performed for LiH and N$_2$ in the STO-3G basis and for H$_2$O in the 6-31G basis. For LiH, we include all 6 molecular orbitals in the active space. For H$_2$O, an active space of 6 electrons in 5 active molecular orbitals, including ($1b_2$, $3a_1$, $3a_1$, $1b_1$, $2b_1$), is used. For N$_2$, the $1s$ and $2s$ orbitals are frozen and then an active space of 6 electrons in 6 molecular orbitals, CAS(6e, 6o) is employed.

\subsection{Benchmark of the ADAFT ansatz}
As discussed in Section~\ref{sec:theory2}, the stationary conditions of the total energy with respect to variational parameters are exactly the same for the $x$-ADAPT and $x$-ADAFT ansatz. Here, we first test the convergence of $x$-ADAPT in comparison with $x$-ADAFT for LiH and H$_2$O molecules. The performance of $x$-ADAPT is numerically assessed by studying the ground-state potential energy curves of the H$_2$O molecule.

\begin{figure}[!htb]
\begin{center}
\includegraphics[width=0.45\textwidth]{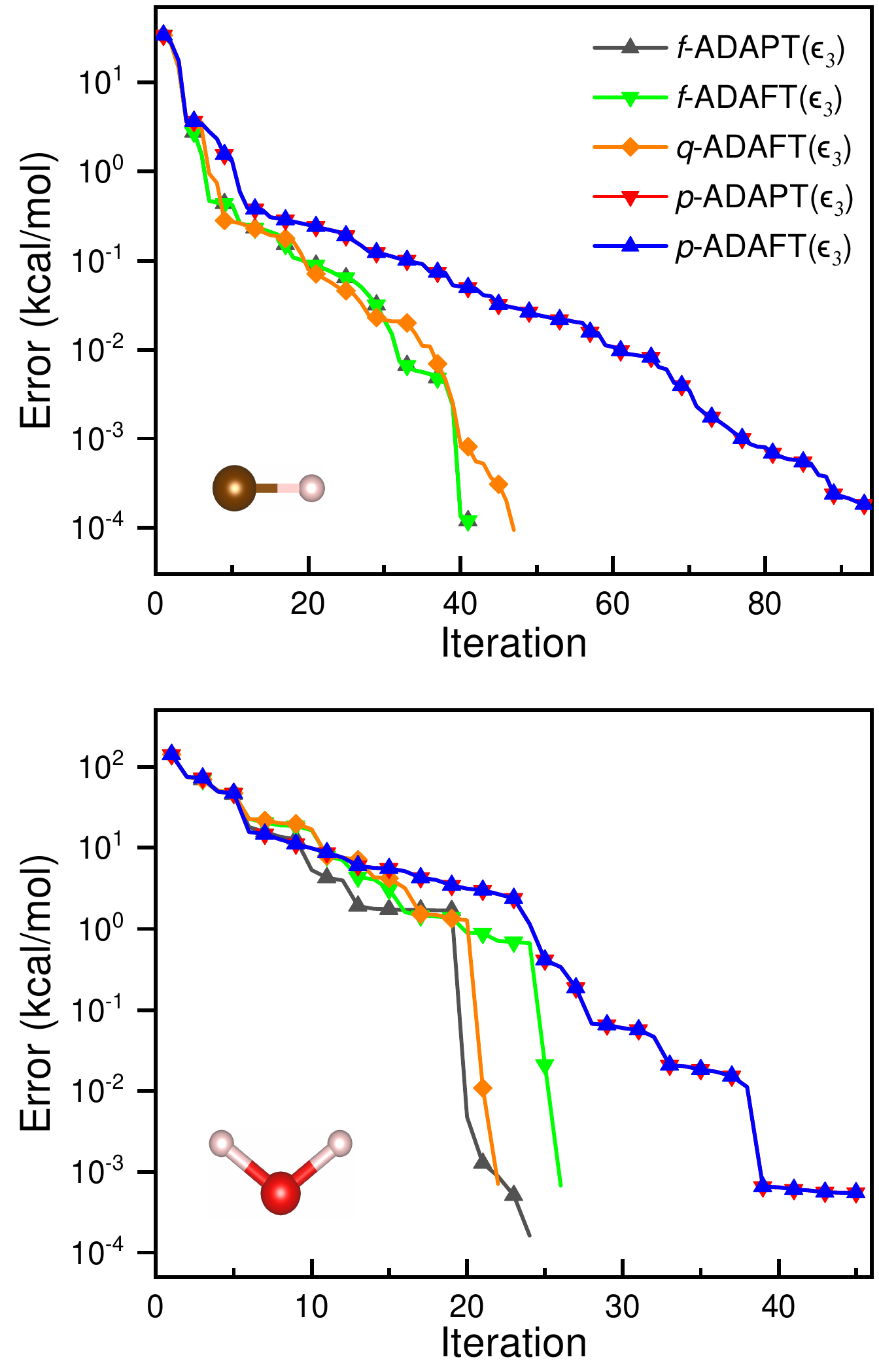}
\end{center}
\caption{Errors of the ground-state energies (in kcal/mol) computed with the ADAPT and ADAFT variational approaches with respect to the exact energy for LiH and H$_2$O. $\epsilon_3$ in the parentheses indicates the convergence threshold $\epsilon=10^{-3}$. The stretched Li-H and O-H bond lengths are 3.0 and 2.5 \AA, respectively.} \label{fig:accuracy}
\end{figure}

\subsubsection{Convergence}\label{sec:conv}
Here, we test the convergence of the $x$-ADAFT ansatz for two dissociated molecular systems that show strong correlation effects: LiH with a stretched bond length of 3.0 Angstrom and H$_2$O with a symmetrically stretched O-H bond length (R$_{\text{O-H}}$) of 2.5 Angstrom and $\angle HOH = 104.5^\circ$.

In figure~\ref{fig:accuracy}, the errors of the ground state energy as a function of the number of iterations are shown for $x$-ADAFT and $x$-ADAPT. It is clear that all adaptive algorithms are able to converge to a given threshold of $\epsilon=1e^{-3}$ Hartree as the number of iterations increases. This demonstrates that, as expected, the $x$-ADAFT ansatz can provide an accurate enough representation of the full configuration interaction wave function with a product of transformations as the $x$-ADAPT ansatz. The rate of convergence for $q$-ADAFT is very close to that of $f$-ADAFT, namely, dropping Pauli-Z strings in Eq.~\ref{eq:operator_qubit} does not significantly influence the convergence. The number of iterations required for $p$-ADAFT to converge to a given accuracy is much larger than that for $f$-ADAFT and $q$-ADAFT since a more general form of Pauli strings is adopted in $p$-ADAFT in the sense that more parameters are necessary to recover the inherent symmetry of the wave function. For LiH, $f$-ADAFT takes 41 iterations, and $p$-ADAFT takes 93 iterations to reach an accuracy of $\epsilon=1e^{-3}$ Hartree. $q$-ADAFT takes 47 iterations to converge to the same accuracy, which is slightly larger than $f$-ADAFT. For H$_2$O, $q$-ADAFT takes almost the same iterations as $f$-ADAFT. In addition, in comparison with $q$-ADAFT and $f$-ADAFT, the error curve of $p$-ADAFT exhibits many plateaus, which results from the fact that the wave function is optimized in a larger parameter space.

One interesting fact revealed in Fig.~\ref{fig:accuracy} is the same energy error curves for $p$-ADAFT and $p$-ADAPT in both cases of LiH and H$_2$O. Considering that operators in the $p$-ADAPT ansatz, its exponential form can be exactly expanded to the first order,
\begin{equation}\label{eq:qubit-u}
    e^{\theta \bar{\tau}_u} = cos(\theta) + sin(\theta) \bar{\tau}_u.
\end{equation}
This is equivalent to the transformation used in the $p$-ADAFT ansatz except for a normalized factor. We are also aware of the same form of unitary operators used in the iQCC approach. Note that the exact expansion of Eq.~\eqref{eq:qubit-u} only works for the case of iteratively updating one operator at a time. When many operators are to be updated, an approximate form, such as first-order Taylor expansion or an involutory linear combination (ILC) from that introduced in Ref.~\citenum{LanRyaIzm21}, should be applied to avoid a finite-order truncation of exponential unitary operators.

\begin{table}[!htb]
    \centering
    \caption{The number of iterations and corresponding variational parameters (in parentheses) of different ADAPT and ADAFT approaches in the ground-state calculation of LiH. $d$ is the number of operators to be updated in each iteration. The convergence threshold is $\epsilon=10^{-3}$ Hartree.}
    \begin{tabular}{|l|ccccc|}
        \hline
       \diagbox[width=6em,trim=l]{method}{d  } & 1 & 5 & 10 & 15 & 20 \\
        \hline
     $f$-ADAFT & 41(41) & 12(60) & 7(70) & 5(75) & 4(80) \\
     $q$-ADAFT & 47(47) & 12(60) & 7(70) & 5(75) & 4(80) \\
     $p$-ADAFT & 93(93) & 21(105) & 14(140) & 9(135) & 8(160) \\
     $f$-ADAPT & 41(41) & 11(55) & 7(70) & 4(60) & 4(80) \\
     $q$-ADAPT & 46(46) & 11(55) & 6(60) & 4(60) & 4(80) \\
     $p$-ADAPT & 93(93) & 20(100) & 13(130) & 9(135) & 8(160) \\ \hline
    \end{tabular}
    \label{table:conv}
\end{table}

Table~\ref{table:conv} shows the number of iterations and corresponding variational parameters for different ADAFT and ADAPT approaches, in which the number of operators to be updated ($d$) in each iteration varies from 1 to 20. As discussed in Ref.~\citenum{LiuLiYan21}, as $d$ increases, the number of iterations at convergence significantly decreases and the number of corresponding variational parameters increases. For example, in the case of $d=20$, $f$-ADAFT converges to $\epsilon=10^{-3}$ in 4 iterations while it takes 41 iterations for $d=1$. The number of variational parameters for $d=20$ is almost twice as many as that for $d=1$. The performance of the ADAFT and ADAPT approaches in convergence is quite similar, while the form of transformations is totally different. In the case of $d > 1$, the ADAPT approach still employs the form of the product of $d$ exponential operators in each iteration, but the ADAFT approach combines $d$ operators together in a linear form, as shown in Eq.~\ref{eq:wave_taylor_n}, which is easy to implement on classical computers.

\begin{figure*}[!htb]
\begin{center}
\includegraphics[width=0.9\textwidth]{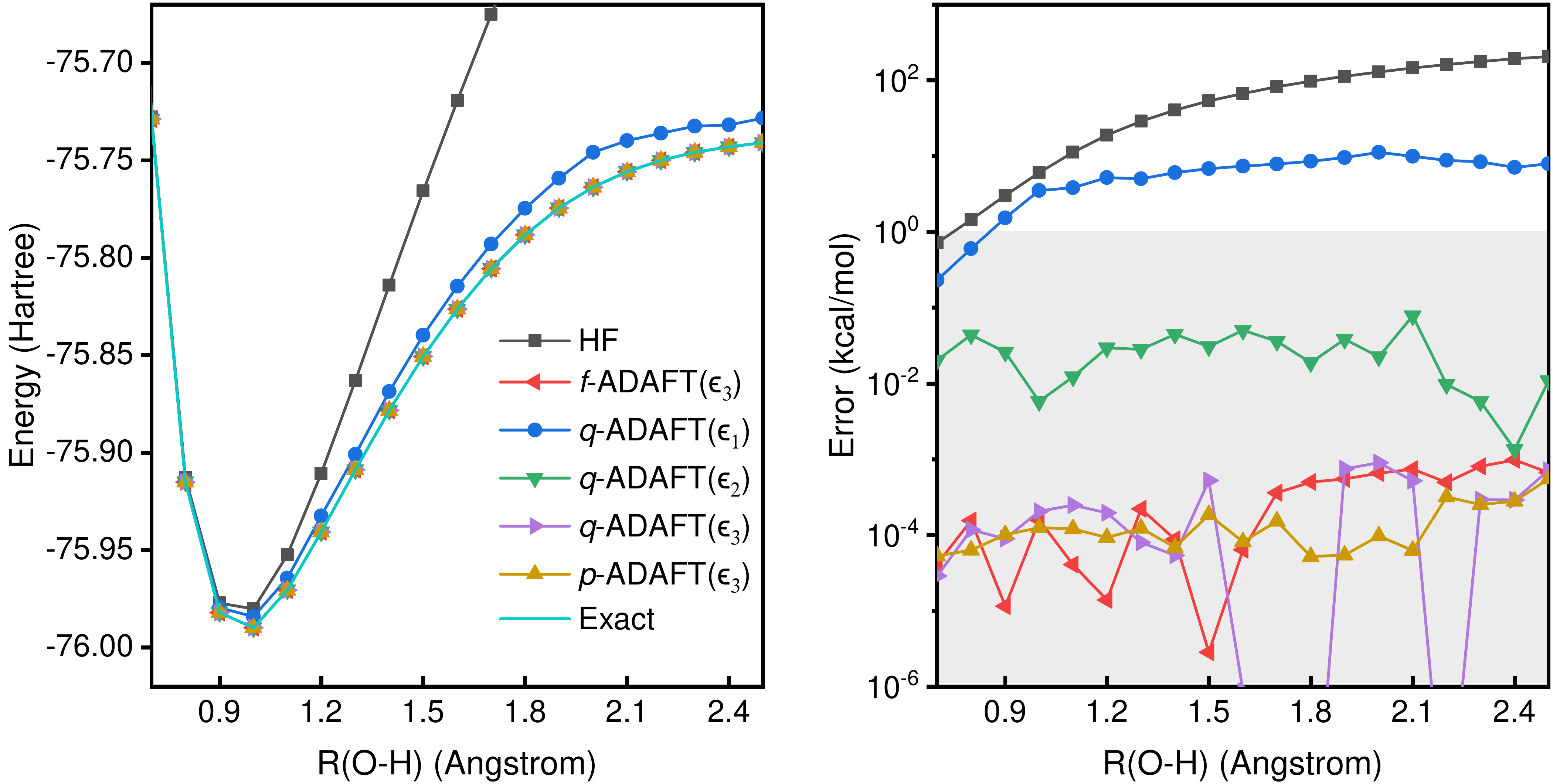}
\end{center}
\caption{The ground-state potential energy curve (in Hartree) and energy error (in kcal/mol) with respect to the exact (CASCI) result for H$_2$O computed with different ADAFT ansatz. $\epsilon_k$ in parentheses indicates the convergence threshold $\epsilon=10^{-k}$. The shaded gray region represents the area within "chemical accuracy" as 1 kcal/mol. The $f$-ADAFT($\epsilon_3$), $f$-ADAFT($\epsilon_3$), and $f$-ADAFT($\epsilon_3$) energy curves lie directly underneath the exact curve, and it is difficult to make a clear distinction.} \label{fig:h2o_taylor}
\end{figure*}

\subsubsection{Accuracy}
The performance of the ADAFT approaches is assessed for the potential energy curves of the symmetric bond stretch of H$_2$O in the 6-31G basis set. For the $q$-ADAFT approach, different values of the threshold with $\epsilon=10^{-1}$, $10^{-2}$ and $10^{-3}$ Hartree are also shown for comparison. Figure~\ref{fig:h2o_taylor} shows the ground-state potential energy curve and the energy errors with respect to the CASCI results for Hartree-Fock (HF), $f$-ADAFT, $q$-ADAFT and $p$-ADAFT. HF provides a very good description of the H$_2$O molecule near the equilibrium bond length, while this situation rapidly deteriorates after symmetric dissociation. The total energy error of HF at an O-H bond length of 2.5 Angstrom is as large as $\sim$200 kcal/mol with respect to the CASCI result.

Overall, $f$-ADAFT, $q$-ADAFT and $p$-ADAFT are accurate enough to describe the H$_2$O symmetric dissociation with a relatively tight convergence threshold. With $\epsilon=10^{-3}$ Hartree, all three methods can achieve a very high accuracy with energy errors less than $10^{-3}$ kcal/mol. There is little difference among $f$-ADAFT, $q$-ADAFT and $p$-ADAFT at a given thresh except for the different numbers of variational parameters, as discussed above. The overall energy errors of $q$-ADAFT($\epsilon_2$) are still within chemical accuracy, while they are 1-2 orders of magnitude larger than those in the case of $\epsilon=10^{-3}$ Hartree. The maximum error of 0.08 kcal/mol in $q$-ADAFT($\epsilon_2$) appears in the case of R$_{\text{O-H}}=2.1$ \AA. In $q$-ADAFT($\epsilon_1$), the maximum error of 11 kcal/mol appears at almost the same place as $q$-ADAFT($\epsilon_2$).

\begin{table}[!htb]
    \centering
    \caption{Nonparallelity error (NPE) (in kcal/mol) for the ground state of H$_2$O computed with the 6-31G basis set. The reference values are from CASCI in an active space of CAS(5o,6e).}
    \begin{tabular}{|c|c|c|c|}
        \hline
            & HF & $f$-ADAPT($\epsilon_3$) & $q$-ADAPT($\epsilon_1$) \\
        \hline
        NPE & 204.91 & 0.00 & 10.60 \\
        \hline
            &$q$-ADAPT($\epsilon_2$) & $q$-ADAPT($\epsilon_3$) & $p$-ADAPT($\epsilon_3$) \\
            \hline
        NPE & 0.08 & 0.00 & 0.00 \\
        \hline
    \end{tabular}
    \label{table:h2o_NPE}
\end{table}

In Table~\ref{table:h2o_NPE}, we present the nonparallelity error (NPE) in the ground state of H$_2$O. Here, NPE refers to the difference between the maximum and minimum error as a useful measure of performance. Although the convergence threshold of $\epsilon=10^{-1}$ Hartree is not accurate enough to describe the double dissociation of H$_2$O with the NPE as large as 10.60 kcal/mol, it significantly improves over HF, especially at the region of a full O-H bond dissociation. The NPE of HF is 204.91 kcal/mol while its error at 0.8 \AA{}  is only 1.44 kcal/mol. As a consequence, $q$-ADAFT($\epsilon_1$) converges in 1 iteration at this small bond length while 12 iterations are required to converge to the same accuracy for R$_{\text{O-H}}=2.3$ \AA.

\begin{figure*}[!htb]
\begin{center}
\includegraphics[width=0.9\textwidth]{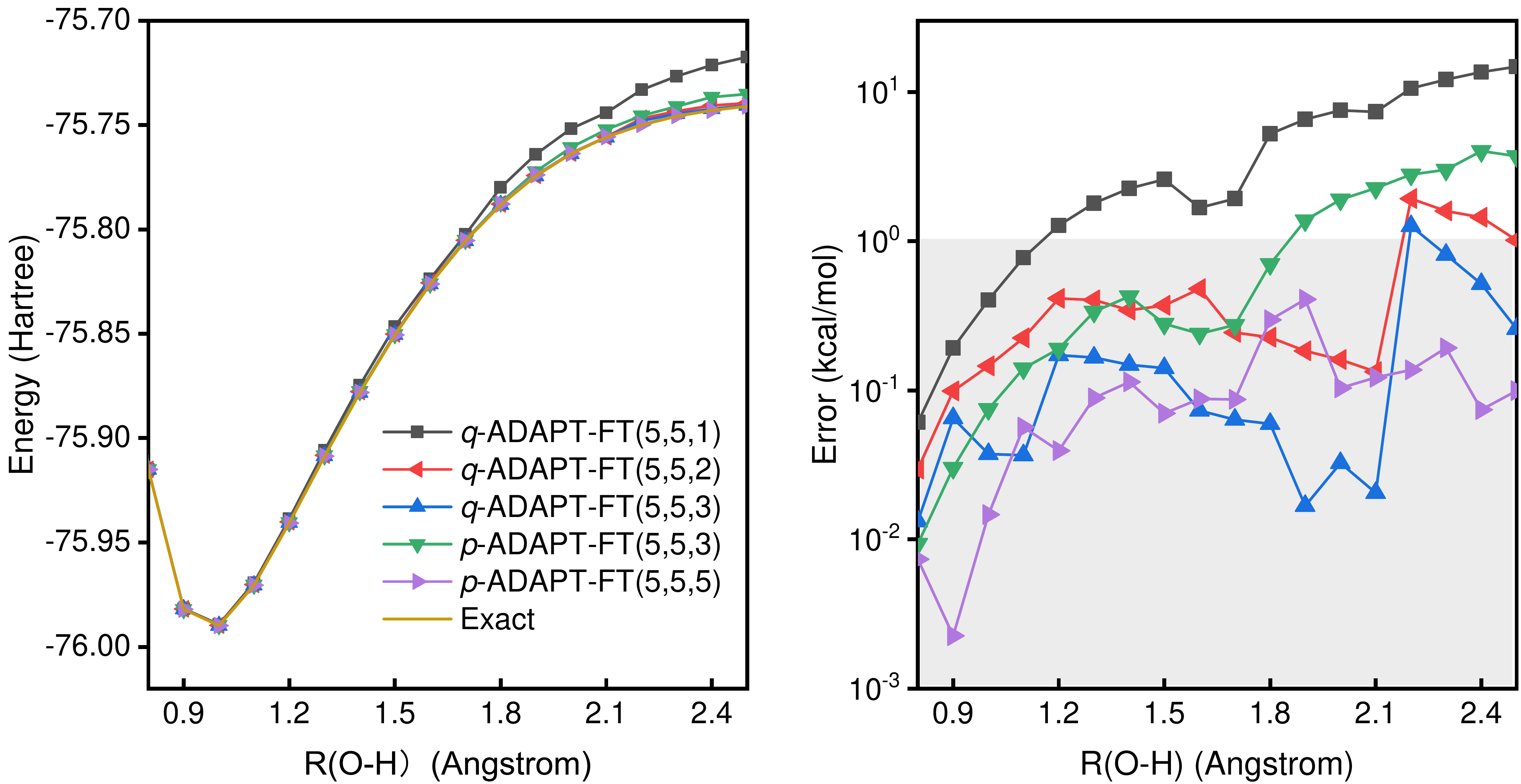}
\end{center}
\caption{The ground-state potential energy curve (in Hartree) and energy error (in kcal/mol) with respect to the CASCI results for H$_2$O computed with the ADAPT-FT($k,d,m$) methods. $k$ is the number of variational parameters in the ADAPT wave function, and $d$ and $m$ are the number of operators updated in each transformation and transformations in the effective Hamiltonian, respectively. The shaded gray region represents the area within "chemical accuracy" as 1 kcal/mol.} \label{fig:h2o_tc}
\end{figure*}

\subsection{Assessment of the Effective Hamiltonian}
As discussed in Sec.~\ref{sec:theory2}, the correlation energy can be partly included in the effective Hamiltonian, and the correlated wave function is prepared with a fixed circuit depth. In this section, we assess the performance of the ADAPT-FT method by varying the number of iterations. Here, $x$-ADAPT-FT($k,d,m$) indicates $k$ variational parameters used in the ADAPT wave function (see Eq.~\ref{eq:wave_adapt}), $d$ operators used to update the transformation in each iteration (see Eq.~\ref{eq:transform_taylor_n}) and $m$ iterations used to construct the effective Hamiltonian. Considering the very similar performance of the use of fermionic and qubit excitation operators, we perform only calculations for the $q$-ADAPT-FT and $p$-ADAPT-FT approaches.

\subsubsection{H$_2$O molecule}
We applied the ADAPT-FT method to study the dissociation potential energy curve of the H$_2$O molecule. In state preparation with the ADAPT ansatz, we fix the depth of quantum circuits, which include a product of five exponential unitary transformations. In the construction of the effective Hamiltonian, we employ five operators in each iteration and consider varying the number of iterations $m$ up to 5 in Fig.~\ref{fig:h2o_tc} to demonstrate the systematic improvability of energy estimations with shallow quantum circuits.

Figure~\ref{fig:h2o_tc} shows the ground state potential energy curve and energy error as a function of O-H bond length for different ADAPT-FT methods. The overall error of $q$-ADAPT-FT(5,5,3) is within chemical accuracy except for the case of R$_{\text{O-H}} = 2.2$ \AA. For the ADAPT-FT methods, the accuracy is improved as the number of iterations $m$ increases while the circuit depth is maintained constant. The mean error of $q$-ADAPT-FT(5,5,m) decreases from 5.04 to 0.22 kcal/mol as $m$ increases from 1 to 3. Although the largest energy error of $q$-ADAPT-FT(5,5,1) ranging from 0.8 to 2.5 \AA  is 14.80 kcal/mol, it is much smaller than the corresponding error of 48.13 kcal/mol in $q$-ADAPT(5). This means that a large part of the correlation energy can be recovered even with a one-step transformation of the Hamiltonian.

\begin{table}[!htb]
    \centering
    \caption{The errors of the total energies (in kcal/mol), the number of CNOTs (n$_C$, the quantum circuit compiled with Qiskit~\cite{qiskit}) and the number of terms in the Hamiltonian (n$_H$) in the calculations of H$_2$O at 2.3 \AA{}. In the $p$-ADAPT($k$) method, $k$ is the number of variational parameters included in the wave function ansatz. In the $p$-ADAPT-FT($k,d,m$) method, $d$ and $m$ are the number of operators and iterations in the transformation of the Hamiltonian.}
    \begin{tabular}{ccccccccc}
    \hline
    \multicolumn{4}{c}{$p$-ADAPT} && \multicolumn{4}{c}{$p$-ADAPT-FT} \\
    $k$ & Error & n$_C$ & n$_H$ &&($k,d,m$)& Error & n$_C$ & n$_H$ \\
    \cline{1-4}\cline{6-9}
    5 & 58.30 & 30 & 292 && &&&\\
    10 & 6.33 & 48 & 292 && (5,5,1) & 6.27 & 30 & 2066 \\
    15 & 4.28 & 78 & 292 && (5,5,2) & 4.41 & 30 & 8944 \\
    20 & 3.87 & 108 & 292 && (5,5,3) & 3.40 & 30 & 23127 \\
    25 & 0.82 & 138 & 292 && (5,5,4) & 0.53 & 30 & 32210 \\
    \end{tabular}
    \label{table:ncnots}
\end{table}

\begin{table}[!htb]
    \centering
    \caption{The number of terms in the original and effective Hamiltonians, labeled N$_i$ and N$_f$, respectively, and the errors of the total energies (in kcal/mol) computed with ADAPT-FT(5,5,1) for H$_2$O at 2.3 \AA{} and ADAPT-FT(8,8,1) for N$_2$ at 2.2 \AA.}
    \begin{tabular}{|c|c|c|c|c|}
    \hline
    Molecule & Methods & N$_i$ & N$_f$ & Errors \\
    \hline
    H$_2$O & $f$-ADAPT-FT(5,5,1) & 292 & 6643 & 4.80 \\
           & $q$-ADAPT-FT(5,5,1) & 292 & 11186 & 13.23 \\
           & $p$-ADAPT-FT(5,5,1) & 292 & 2066 & 6.27 \\
        \hline
    N$_2$  & $f$-ADAPT-FT(8,8,1) & 307 & 55035 & 35.28 \\
           & $q$-ADAPT-FT(8,8,1) & 307 & 62820 & 39.31 \\
           & $p$-ADAPT-FT(8,8,1) & 307 & 5120 & 45.51 \\
        \hline
    \end{tabular}
    \label{table:nterms}
\end{table}

In comparison with $q$-ADAPT-FT, $p$-ADAPT-FT converges slowly as $m$ increases. For example, the mean error of $q$-ADAPT-FT(5,5,3) is 0.22 kcal/mol, much smaller than the 1.21 kcal/mol of $p$-ADAPT-FT(5,5,3). This result is consistent with the rate convergence of the ADAFT approach, as discussed in Sec.~\ref{sec:conv}. When $m$ increases to 5, the $p$-ADAPT-FT(5,5,5) method also provides an accurate description of the double dissociation of H$_2$O. It is worth mentioning that the mean errors of $q$-ADAPT-FT(5,5,3) and $p$-ADAPT-FT(5,5,5) are very close to each other. Nevertheless, $p$-ADAPT-FT(5,5,5) performs more stably, while there is a sudden jump in $q$-ADAPT-FT(5,5,3) at 2.2 \AA. This situation often occurs in adaptive quantum algorithms when the wave function ansatz is far from the exact one.~\cite{GriEcoBar19}

In principle, the ADAPT-FT method reduces the quantum circuit depth at the expense of the increasing number of terms in the effective Hamiltonian. Table~\ref{table:ncnots} shows the number of two-qubit CNOT gates and the number of terms in the Hamiltonian in a variety of $p$-ADAPT(k) and $p$-ADAPT-FT methods. In the $p$-ADAPT(k) methods, the errors of the ground state energy decrease from 58.30 to 0.82 kcal/mol as the number of variational parameters $k$ increases from 5 to 25. Correspondingly, the number of CNOTs required for implementing the wave function ansatz increases from 30 to 138 and the number of terms in the Hamiltonian remains a constant of 292. In contrast, the number of CNOTs remains constant in the $p$-ADAPT-FT(5,5,m) methods, and the number of terms in the effective Hamiltonian increases from 2066 to 32210 as $m$ increases from 1 to 4. As expected, the errors of the ground state energy are comparable when the number of operators involved in the ADAPT and ADAPT-FT methods is the same. For example, the $p$-ADAPT-FT(5,5,4) method updates the wave function ansatz with 5 operators and constructs the effective Hamiltonian with 20 operators, and its error of 0.53 kcal/mol is close to the energy error of $p$-ADAPT(25).

The electronic Hamiltonian for water with 5 active orbitals contains 292 terms. The effective Hamiltonian generated from a transformation that is composed of a linear combination of 5 operators contains 6643, 11186, and 2066 terms in $f$-ADAPT-FT, $q$-ADAPT-FT, and $p$-ADAPT-FT (see Table~\ref{table:nterms}). It is clear that $p$-ADAPT-FT(5,5,1) has a much slower growth factor of 7 in comparison with the growth factor of 23 in $f$-ADAPT-FT(5,5,1) and 38 in $q$-ADAPT-FT(5,5,1). The error of $p$-ADAPT-FT(5,5,1) is only 6.27 kcal/mol, slightly larger than 4.80 kcal/mol in $p$-ADAPT-FT(5,5,1) but smaller than 13.23 kcal/mol in $q$-ADAPT-FT(5,5,1). This is consistent with the fact revealed in Fig.~\ref{fig:accuracy} that $p$-ADAPT and $p$-ADAFT show a similar convergence with $f(q)$-ADAPT and $f(q)$-ADAFT at the very beginning of iterations.

\subsubsection{N$_2$ molecule}
We also apply the ADAPT-FT method to study the N-N bond dissociation of N$_2$, which is a very challenging system for traditional electronic structure methods since it involves the breaking of a triple bond~\cite{Sie83,MaLiLi06}. Typically, there are a total of 6 electrons that are strongly entangled when the N-N bond elongates. In the construction of the effective Hamiltonian, we use eight operators to build the linear transformation and consider varying the number of iterations up to 5 for $q$-ADAPT-FT and 10 for $p$-ADAPT-FT. The number of iterations for state preparation is fixed to 6.

\begin{figure*}[!htb]
\begin{center}
\includegraphics[width=0.9\textwidth]{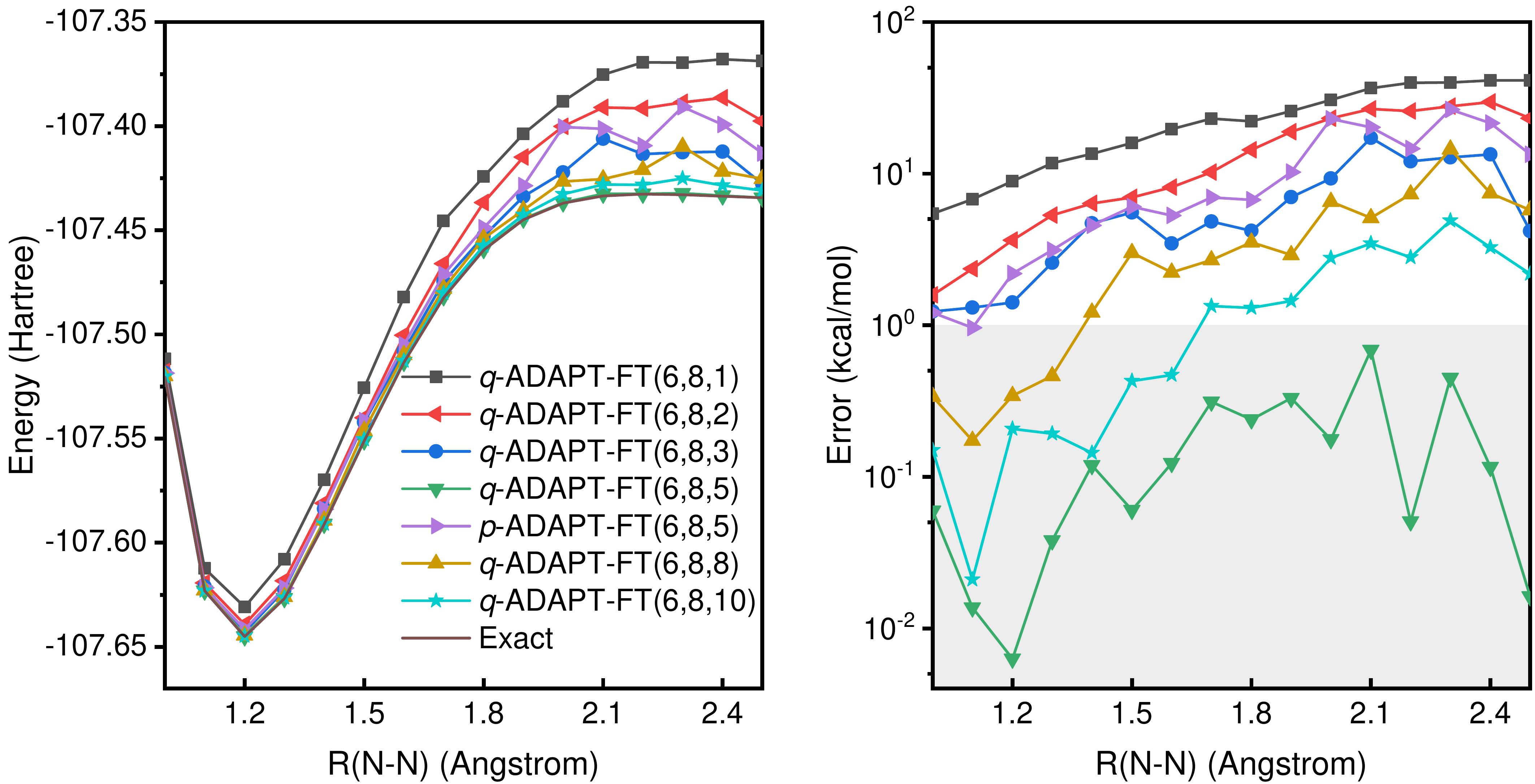}
\end{center}
\caption{The ground-state potential energy curve (in Hartree) and energy error (in kcal/mol) with respect to the CASCI result for N$_2$ computed with the ADAPT-FT($k,d,m$) methods. The shaded gray region represents the area within "chemical accuracy" as 1 kcal/mol.} \label{fig:n2_tc}
\end{figure*}

Figure~\ref{fig:n2_tc} shows the ground state potential energy curve and the absolute energy error as a function of the N-N bond length for different ADAPT-FT methods. The overall error of $q$-ADAPT-FT(6,8,5) is less than 1 kcal/mol. Compared with H$_2$O, the energy error with respect to the number of iterations $m$ converges much slower for N$_2$ so that larger parameters of $(k,d,m)$ are employed in N$_2$ calculations. The mean error of $q$-ADAPT-FT(6,8,m) decreases from 23.87 to 0.17 kcal/mol as $m$ increases from 1 to 5. However, the mean error of $q$-ADAPT-FT(6,8,m) decreases from 10.40 to 1.57 kcal/mol as $m$ increases from 5 to 10. Therefore, even with $m$ as large as 10, $p$-ADAPT-FT(6,8,m) is unable to achieve chemical accuracy. On the other hand, quantum circuits with 6 variational parameters in $p$-ADAPT-FT are in principle shallower than those in $q$-ADAPT-FT. As such, for a complex system, it is uncertain whether we can establish the advantage of $p$-ADAPT-FT over $q$-ADAPT-FT even though a much simpler form of operators is used in $p$-ADAPT-FT.

When $m$ is not large enough, remarkable fluctuations exist in the potential energy curve of $q$-ADAPT-FT and $p$-ADAPT-FT as shown in Figure~\ref{fig:n2_tc}. In adaptive quantum algorithms, the effective Hamiltonian is self-consistently grown. Therefore, when the number of variational parameters in the ADAPT-FT method is fixed, a discontinuity may appear if the energy does not converge. For example, there exists a sudden drop in the energy curve at 2.2 \AA.

The initial Hamiltonian of N$_2$ with 6 active orbitals contains 307 terms. In comparison with the water molecule, the inclusion of 8 operators in the transformation results in a much larger growth factor in the effective Hamiltonian. For example, the growth factor of $p$-ADAPT-FT for N$_2$ is 17, which is $\sim$2.5 times that for H$_2$O. This is consistent with the growth scaling of $\mathcal{O}(d^2)$ in the effective Hamiltonian in the form of Eq.~\ref{eq:dressed_hamiltonian}. The errors of the $f(q,p)$-ADAPT-FT(8,8,1) approaches are 35.28, 39.31 and 45.51 kcal/mol, respectively, which exhibit little difference among these three approaches. As a result, the $p$-ADAPT-FT approach is a good candidate to estimate the exact energy using shallow quantum circuits.

\section{Conclusion}\label{sec:conclusion}
In this work, we propose an adaptive electronic structure approach, in which the exact wave function is approximated as the product of a series of transformations in the form of a linear combination of many operators acting on the reference wave function. This approach can be regarded as a first-order Taylor expansion of the ADAPT ansatz so that we name it as the ADAFT approach. The ADAFT approach is expected to be able to achieve the same accuracy as the ADAPT approach and it has been numerically demonstrated in potential energy curve calculations of small molecules, LiH and H$_2$O.

Because the transformation in the ADAFT approach is not unitary, this approach is not suitable for direct implementation on quantum computers. However, considering the linear combination form of the transformations, it provides a promising scheme to construct the effective Hamiltonian. We further combine it with the ADAPT approach and propose a resource-efficient quantum computational chemistry method, the ADAPT-FT algorithm. We assess the performance of the ADAPT-FT algorithms by numerically studying the dissociation potential energy curve of H$_2$O and N$_2$ molecules and demonstrate the systematic improvability of energy estimations with fixed-depth quantum circuits.

Similar to the iQCC, the accuracy of ADAPT-FT is improved by increasing the number of transformations used to construct the effective Hamiltonian. However, this leads to a fast growth of the number of terms in the effective Hamiltonian (namely, rapidly increasing measurements) as discussed in iQCC. Therefore, several advanced techniques, e.g., compression and extrapolation technologies, introduced in iQCC can be used to improve the $p$-ADAPT-FT.

\section{Acknowledgments}
This work is supported by the National Natural Science Foundation of China (22073086 and 21825302,21688102).



\vspace{3ex}

\end{document}